# Close Communication and 2-Clubs in Corporate Networks: Europe 2010.

Robert J. Mokken and Steven Laan *

*Abstract*: Corporate networks, as induced by interlocking directorates between corporations, provide structures of personal communication at the level of their boards. This paper studies such networks from a perspective of close communication in sub-networks, where each pair of nodes (boards of a corporation) are either neighbours, or have a common neighbour. These correspond to subgraphs of diameter at most 2, designated by us earlier as 2-clubs, with three types (coteries, social circles and hamlets) as degrees of close communication in social networks, within the concept of boroughs of a network. Boroughs are maximal areas and containers of close communication between nodes of a network. This framework is applied in this paper to an analysis of corporate board interlocks between the top 300 European corporations 2010, as studied by Heemkerk (2013), with data provided by him for that purpose. The paper gives results for several perspectives of close communication in the European corporate network of 2010, a year close to the global crash of 2008, as a further elaboration of those given in Heemskerk (2013).

* University of Amsterdam, Informatics Institute. Corresponding author: mokken@uva.nl

# 1 Introduction

Heemskerk [2] gave an elaborate analysis of the network of interlocking directorates of the major European corporations in 2005 and 2010.

In such corporate networks interlocks provide channels of personal access and communication between the boards of corporations, so that areas of close communication can be defined as sub-networks where each pair of nodes (corporations) are neighbours or have at least one common neighbour. These can be represented as (sub)graphs of diameter at most two. For such sub-networks Mokken [3-5] introduced the concept of 2-clubs of a network and its three types (coterie, social circle, hamlet), later extended by Laan *et al* [1] with the concept of boroughs, which contain the 2-clubs of a network. Recent advances in hardware and corresponding programming techniques provide means and opportunities to use them on large networks (*e.g.* [6; 7] We tried to apply these in the context of corporate networks, using software developed by one of us [8; 9] . From this perspective of close communication we extend Heemskerk's analysis of the major European corporations in 2010, using his data set. In the following sections we first introduce the conceptual and analytic framework. We then analyse Heemskerk's network of 286 major European companies, restricting ourselves to its major component of 259 firms.

# 2 Conceptual framework[1]

*Close communication* in a network is defined here as access and communication between nodes directly between neighbours ($1^{st}$ step) or through a common neighbour ($2^{nd}$ step). *Close communities* in a network are areas where each pair of nodes are neighbours or have a common neighbour. These can be represented as graphs of diameter at most two. Conform [3; 4] we represent *close communities* in a network by its 2-clubs, which are *maximal* subgraphs of diameter at most two: they are not included in, or part of another subgraph of diameter at most two. Note that 2-clubs can and will overlap heavily in the sense of having many nodes and edges in common.

In [5] we showed showed that there are just three types of 2-clubs, or close communities, conform three levels of close communication.

## 2.1 Types of 2-clubs

The *first* type (c*oterie*) is related to the ego-networks of the nodes of a network. For each (non-isolated) node of a network its *ego-network* is the subgraph of that node together with its neighbours and all edges joining them, with the node as single central point: its centre or ego. Obviously any ego-network has diameter at least 2, but it is only a 2-club, a c*oterie,* if it is not included or part of any other 2-club of the network. Each coterie has a central node, which identifies it. Coteries are tightly meshed, involving communication along triangles only, thus confining their level of close communication to strictly local, *within* the ego-network around their central point or ego. (An example is given in Figure 1)

The other two types or close communities (2-clubs) concern close communication *between* ego-networks of a network.

---

1   We assume familiarity with the most basic graph and network notions *e.g.* [10]. For a more complete account see Laan et al [1] .

The *second* type (s*ocial circle*) is a 2-club without a central point, but with at least one central pair of neighbours *i.e.* each node of a social circle is neighbour of at least one of the two nodes of that central pair. We shall see an example in Figure 3.  Social circles, constituted by triangles and squares, are more loosely meshed. Their level of close communication is confined to an intermediate local level *between* the ego-networks of their central pairs of neighbours. The minimum size of a social circle is 4: a rectangle or cycle of length 4 (C4)

Finally, the *third* type (*hamlet*) is a 2-club without central node or central pairs. Hamlets are constituted by triangles, squares, and pentagons, and thus are more widely meshed.  Without central nodes or central pairs, they involve close communication between (parts of) ego-networks at the widest local level. We shall see an example in Figure (6). The minimum size for a hamlet is 5: a cycle of length 5 (C5) or pentagon.

Thus social circles and hamlets involve close communication *between* ego-networks, whereas coteries, as ego-networks, just concern communication *within* these ego-networks around their central node (ego). Our main focus will therefore be on the social circles and hamlets of a social network.

## 2.2  Boroughs

The 2-clubs of a connected network are located in its boroughs ([1] ). A *borough* of a connected network is a *maximal* subgraph with the property that each edge is on a basic cycle (triangle, square, or pentagon) of the network, and therefore also part of one or more 2-clubs of the network. Consequently, the 2-clubs of a network are located in just one of the boroughs, which roughly can be seen as a collection of overlapping and edge chained 2-clubs. Thus, for practical purposes, boroughs can be seen as disjoint collections of all the 2-clubs of a network, forming delimited dense areas of close communication in that network.

# 3  European corporate network  2010

Heemskerk [2] gave an extensive comparative analysis of the network of largest stock listed European companies in 2005 and 2010,  as listed in FTSE Eurofirst top 300 index,  focussed on the development of this European network between 2005 and 2010 as an economic institutional network in a period where the political European Union had to cope with the effects of the financial crisis of 2008. Although not ignoring these objectives, in this paper we analyse Heemskerk's  dataset for 2010 solely with the purpose to delve deeper in the close communication areas of this real network and experiment with the associated concepts and methods.[2] For background and details we refer to [2].

The detection of all 2-clubs in a large and dense network will result in a multitude, if not myriad, of mutually heavily overlapping 2-clubs of the three types. To detect these we used a two-step approach: first finding the boroughs in the components of a network, as the containers of its non-trivial 2-clubs, and then in a second step for each, or selected boroughs to detect and store the 2-clubs contained in them ([1] , [8], [9]).

Our prototype software stores the 2-clubs in a database, according to type.  One of us [8; 9] developed a preliminary tool to investigate, search and analyse this database, which enabled us to derive some results as reported in the following sections.[3]

---

2   We thank Eelke Heemskerk for making this dataset available to us.
3   A first beta-version of the software is available at   https://github.com/Neojume/TwoClubs  [9]

## 3.1 The European borough 2010

We analysed the data as a simple graph, where the nodes represented firms and a single edge joined two nodes if the firms had at least one interlock, *i.e.* one or more common directors. The network covered 286 of the major firms and contained a dominant component (maximal biconnected subgraph) of 259 firms: each pair of nodes in the component is joined by a path in the network.

In this component we found a single dominant borough of 225 firms, in addition to three smaller, trivial boroughs. These coincided with single and disjoint minimal 2-clubs: one of four firms (one British, two Italian, and one Spanish) and two of three firms (one of 3 Swiss firms and one of two Spanish and a Portuguese firm).

We confined further analysis to this giant European borough, covering 79% of all firms in the network and 87 % of those in the dominant component. It is the delimited maximal area of close communities within the European corporate network of 2010 and its major component. It is packed with (almost) all of its sequentially edge-chained 2-clubs, so that each edge is on at least one 2-club. Yet as an area in the network it is rather widely stretched, as its diameter (the longest distance between two of its nodes) is 7.
This European borough contained a total 2128 2-clubs of size 4 or more, distributed over the three types as given in Table 1.

| Type of 2-Clubs Borough | Coteries | Social Circle | Hamlet | Total |
|---|---|---|---|---|
| *# 2-Clubs* | 138 | 717 | 1273 | 2128 |
| *Coverage nodes Borough* * | 99.6% | 89.4% | 92.5% | 100.0% |

*Table 1: European borough 2010: number and type of 2-clubs, size at least 4.*

*\* Coverage: % of nodes of Borough by nodes in type 2-club*

## 3.2 Coteries

The ego-network of 138 companies, 61% of the 225 companies in the borough, formed a coterie of the European borough, because they were not part of any larger 2-Club of the borough. Together these 138 ego-networks cover about all (99.6%) of the 225 companies of the borough.

The ego-networks of the 97 (39%) other firms were included in one or more other 2-clubs.
For instance, the ego-network of *Volkswagen AG* was not a coterie but, with degree 12, included in two other 2-clubs of the borough, both hamlets; while *Deutsche Bank AG*, with degree 8, was included in a German social cycle, where it formed one of its two central pairs from *Bayer AG* (the other *Deutsche Post AG*).
As to size (the number of their firms/nodes) coteries tended to be smaller than the two other types of 2-clubs: the coteries had median size 9, against a median size of 14 for social circles and 16 for hamlets. For instance, ranked according to size, the 83 2-clubs of sizes 22 or larger counted only four coteries, among 35 social circles and 44 hamlets.

Yet, the largest 2-club in the borough was the French coterie of size 27 with *GDF Suez SA* as its centre (See Figure 1). In this ego-network *GDF Suez SA* (nowadays *Engie SA*), a French electricity and gas multinational, had 26 neighbours: 16 French, two British, three Belgian, one German, one Italian, one Dutch, one Luxembourg, and one Swiss companies. Hence it was predominantly French regional (together with the francophone Belgian and Luxembourg companies).

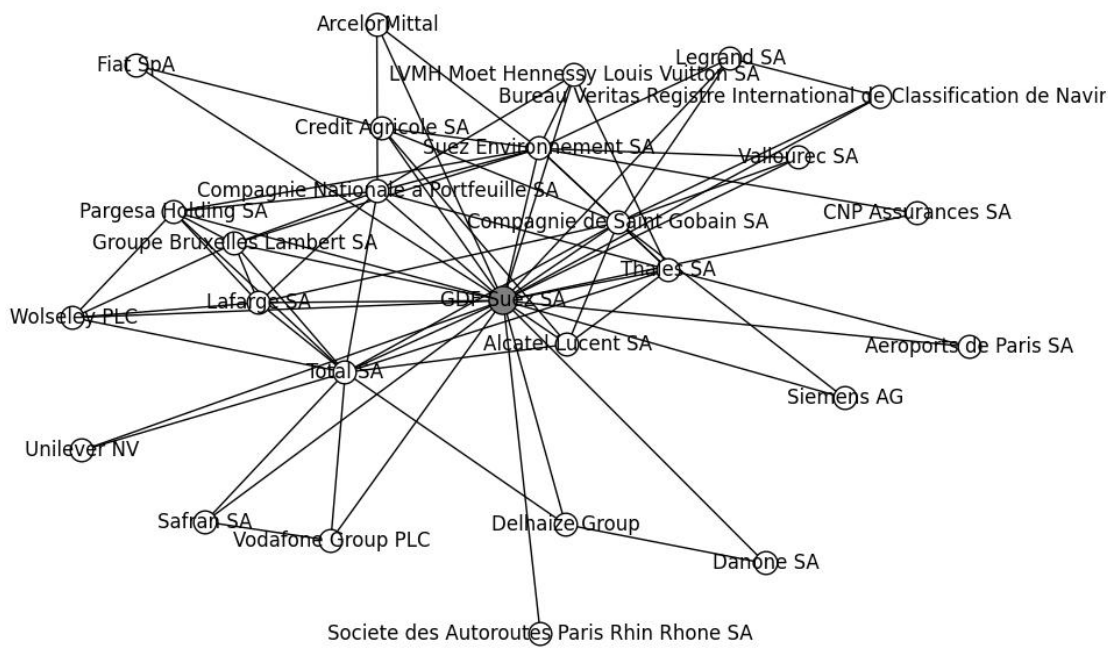

*Figure 1: Coterie of GDF Suez SA: 27 firms of which 21 French regional*

In this paper we define the *scope* of a 2-club as the total number (or percentage) of 2-clubs of the borough, each of which shares at least one node with that 2-club. Similarly, we define the scope of a set of firms by the total number (or percentage) of 2-clubs in the borough that share at least one node with that set.
Thus the ego-enter of the French company *GDF Suez SA*, both the largest 2-club and coterie of the Borough, had a scope of 1804 of the 2-clubs of the European borough 2010, that is 84.8% of its 2-clubs shared one or more firms in the coterie (ego-centre) of *GDF Suez SA*.
In fact the six largest coteries, according to their central firm, were French:

*GDF Suez SA* (size 27);
*Total SA* (size 25);
*Sanofi Aventis SA* (size 22);
*Compagnie de Saint Gobain SA* (size 22);
*LVMH Moet Hennesy Louis Vuitton SA* (size 21);
*Thales SA* (size 21).

The data for the coteries, as identified by their central firm, show a correspondence with Heemskerk's [2] findings for the firms according to size and composition[4]. So, as stated above, we shall confine our interest to those close community structures in the European borough 2010, which involve close communication *between* ego-centres: its social circles and hamlets.

## 3.3 Social circles

As introduced above, social circles involve the intermediate level of close communication between ego-centres. They are narrowly meshed sub-networks of communication along basic triangles and rectangles. They don't have a single central point, but one ore more central pairs of neighbours instead, so that, for each central pair, all other nodes are adjacent to (neighbour of) at least one of its nodes.

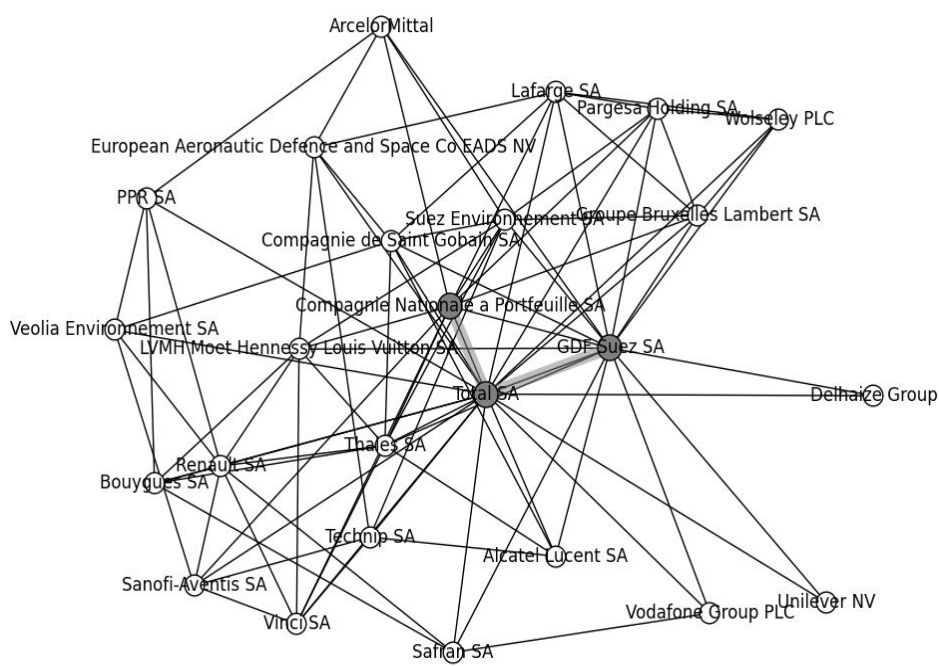

*Figure 2: French social circle with two central pairs from Total SA*

There were 717 social circles in the borough (see Table 1), their nodes (firms) covering 89.4% of its firms. The four largest social circles had size 25 and were French, as they were all spanned by one up to seven central pairs, each involving the pair of *GDF Suez SA* and *Total SA*, the French oil and Gas company, and centre of the second largest coterie in the borough, supplemented with central pairs from *Total SA* with other French companies.

For instance, Figure 2 illustrates one of these with two central pairs from *Total SA*: one with *GDF Suez SA* and the other with the *Compagnie Nationale à Portefeuille SA*, up to 2011 a francophone Belgian investment holding of the Frère family, which was shown by [[1]] to be embedded in the French regional network in close association with *BNP Paribas SA*.

---

4   Note that Heemskerk (2013) counts interlocks between firms, whereas in our network we count neighbours: multiple interlocks between two firms are represented by a single edge.

All 35 social circles with sizes larger than 21 were French, in the sense that al their central pairs were formed by the centres of the largest French coteries.

The largest non-French social circle is a German one of size 21 with central pair *Man SE-Deutsche Telekom AG*. (See Figure 3). Next to its 17 German firms it has three Swiss firms (*Schindler Holding AG*, *ABB Ltd*, *Novartis AG*) and one Swedish firm (*Scania AB*).

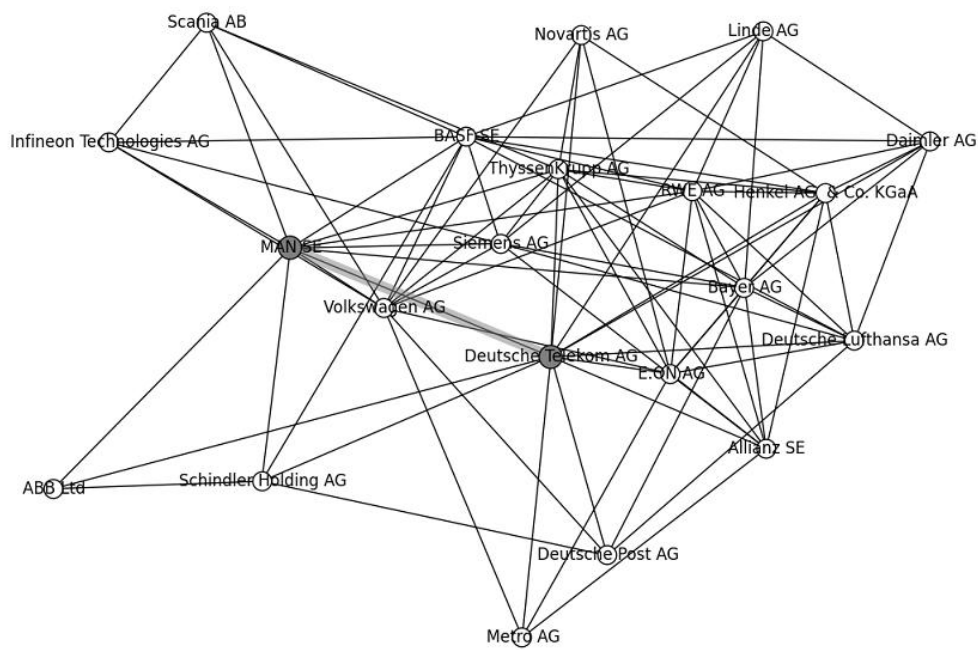

*Figure 3: Largest German social circle, 21 nodes: 17 German, 3 Swiss and 1 Swedish*

## 3.4 Hamlets

Hamlets contains close communication at its widest level: widely meshed networks of basic triangles, rectangles or pentagons without a central node or central pair of nodes. The borough contained 1273 hamlets, the nodes of which covered 92.5 % of those of the borough (See Table 1).

The five largest hamlets had size 24. Given the size of their pairwise common overlap (22-23) these represented just two different types, with similar composition. Each type consisted predominantly of the central companies from the French major coteries, and a few non-French companies, e.g. two Belgian-French Financials (*Groupe Bruxelles Lambert SA* and *Compagnie Nationale a Portfeuille SA*), two British companies (*AstraZeneca PLC*, *Wolseley PLC*, or *Vodafone Group PLC*) and a Swiss financial company (*Pargesa Holding SA*).

Moreover, and similarly, all 44 hamlets of size 22 or larger were predominantly French.
The first largest non-French hamlets are three strongly similar German hamlets of size 21 with large overlaps of 19-20 common firms. The largest predominantly British hamlets sere found with sizes 11 or 10.

# 4   National regions and pivotal 2-clubs

As can be seen from the coverage percentages in Table 1, 2-clubs overlap heavily within and between the three types of 2-clubs. Moreover, the analysis in the previous section suggested that the French sphere of close communication appeared rather dominant and dense in the European borough.
In order to focus and sort out other regional spheres of close communication in the borough, we introduce the idea of a common *pivotal* 2-club, or *pivot,* for an adaptively selected set of firms of a certain category, *e.g.* country, region, or industry.

## 4.1   Procedure of pivot selection

This procedure consisted of an adaptive cumulative selection of firms from a country, according to the size of their coterie or rank in The Global 2000 – Forbes.com [10], starting with the largest company and downward as long as their set of common 2-clubs is not empty. The final set of common 2-clubs thus is not always unique, containing not one, but possibly a few 2-clubs of different, but usually heavily overlapping 2-clubs. From these a 2-club was designated as a pivotal 2-club or pivot for that set, considered as most representative for that set of firms. Focussing at the widest level of close communication, we chose for a pivot the largest common hamlet. If not present then the largest hamlet with complete overlap with the largest social circle, else that social circle itself.
In the following subsections we recapitulate shortly per country or region the results of this procedure, as summarized in Table 2, while referring to pictures of the relevant pivots in the Appendix.

## 4.2   Regional pivots

Heemskerk's dataset for 2010 covered 16 European countries [2]. Our European borough for 2010, derived from Heenskerk, did not contain any major company from Portugal or Greece. The three firms from Luxembourg were all part of French regional networks, as the single Irish firm was part of the British regional networks. Hence the results summarized in Table 2 refer to 12 European countries: 10 from the European Union and two from non-EU countries.

*European Union*

**France.** Twelve French companies and the bank *BNP Paribas SA* (the centres of the thirteen largest French coteries) had one regionally homogeneous hamlet of size 22 in common, consisting of 21 French companies and one Belgian-French company ( *Compagnie Nationale a Portfeuille SA*) .
We designated this hamlet as a pivot for the francophone region in the borough (see Figure 4). The *scope* of that French pivot in the European borough, as defined before (see Section 3.2, p. 5) , was 1618 2-clubs of the borough, *i.e.* 76.0% of the 2128 2-clubs of the borough had one or more firms in that French pivot as well.

| REGION | TYPE PIVOT | SIZE | SCOPE * | |
|---|---|---|---|---|
| *EU* | | | *Total* | *% of Borough* |
| France | Hamlet | 22 | 1618 | 76.0% |
| Germany | Hamlet | 21 | 627 | 29.5% |
| United Kingdom (UK) | Hamlet | 11 | 1724 | 81.0% |
| Italy | Hamlet | 10 | 791 | 37.2% |
| Spain | Hamlet | 6 | 72 | 3.4% |
| Belgium | Hamlet | 14 | 1521 | 71.5% |
| Netherlands (NL) | Hamlet | 12 | 556 | 26.1% |
| Sweden | Social circle | 15 | 419 | 19.7% |
| Denmark | Hamlet | 7 | 277 | 13.0% |
| Finland | Hamlet | 8 | 449 | 21.1% |
| *Non EU* | | | | |
| Norway | Hamlet | 7 | 33 | 1.6% |
| Switzerland | Hamlet | 8 | 296 | 22.5% |

*Table 2: Regional pivotal 2-clubs*

*\* Scope: number or percentage of 2-Clubs of Borough containing at least one firm from pivot*

**Germany.** Twelve German firms, centres of the twelve largest German coteries (sizes 12 to 18), were all part of just two, almost completely overlapping 2-clubs, both hamlets, each of size 21. One of these was chosen as a representative pivot, for the central part of the German regional network, containing 21 companies, including the largest German bank *Deutsche Bank AG,* and all German companies, except for one Swiss healthcare company *Novartis AG* (see Figure 5).

For the 21 firms of this German pivot Table 2 shows a scope of 627 2-clubs or 29.5% of the 2-clubs of the borough: 627 2-clubs of the borough had also at least one firm of this German pivot.

Thus we see that the European borough 2010 contained also a German regional sub-network of close communication, but with a scope of less than half of that of the French regional sub-network.

**United Kingdom.** The central companies of the four largest British coteries (sizes 13 – 20) shared only three 2-clubs, 2 hamlets and a social circle with a mixed Dutch-British central pair *Unilever NV - Smith & Nephew PLC*. Taking into account the heavy overlaps we selected the largest hamlet of size 11 (the four largest British and eight other companies) as pivotal for the central part of the British regional network (See Figure 6).

Its 11 companies (7 British, 2 French, 1 Dutch and 1 Swedish), cover 1724 2-clubs of the European corporate borough: each of those 2-clubs has at least one company of that hamlet as a member.

Together these 1724 2-clubs form 81.0% of the 2128 2-clubs of the borough. Note that no British bank was part of that pivot.
Thus, its mixed composition, with four central companies from France (2), The Netherlands (1), and Sweden (1) extends its access to 81% of the 2-clubs of the European borough, even more than the French pivot (See Table 2).

*Italy*. Five of the largest Italian firms, including the bank *Mediobanca - Banca di Credito Finanziario SpA*, shared one 2-club, a hamlet of size 10. Apart from these five companies it contained another four Italian firms and one French firm ( *Veolia Environnement SA*). Hence this hamlet of nine Italian companies and one French can be considered as a pivot for the Italian regional corporate network (See Figure 7).
Together Its 10 companies had a scope of 791 2-clubs of the European corporate borough: each of those 2-clubs has at least one company of that hamlet as a member. Together these 791 2-clubs form 37.2%, of the 2128 2-clubs of the borough (See Table 2).
Thus its scope, although larger than that of the German pivot was less than half that of France.

*Spain.* The area of close communication for this large member of the European Union is spectacularly low. A single common, elementary hamlet of size 6 was found, which proved to be mixed bi-national Southern-European, consisting of three Spanish and three Italian companies (See Figure 8).
Its six companies have a scope of only 72 2-clubs, or 3.4% of the European corporate borough, suggesting a marginal, if not separated position of Spanish regional companies in the close communication areas of that borough (See Table 2).

*Belgium*. Five Belgian companies shared just two 2-clubs: a coterie of size 13 and a social circle of size 9. The social circle (size 9) had central pairs from Belgacom with, respectively, *Alcatel Lucent SA, Thales SA, Companie Nationale à Portefeuille SA* en *Delhaize Group*) consisting of five Belgian and four French companies. This social circle was maximally included (but for one firm) in one hamlet of size 14 which we selected as pivotal 2-club for the Belgian region. This pivot consisted of 10 French and four Belgian companies, confirming that this region is mainly part of the larger French regional network (See Figure 9).
Because of that francophone orientation we can see from Table 2 a scope of 71.5% of the borough: its 9 companies cover 1360 2-clubs of the European corporate borough: each of those 2-clubs has at least one company of that Belgian hamlet as a member. This scope is more than twice that of the German pivot, and indicates that the Belgian pivot shares a major part of that of the French pivot.

*The Netherlands*. Nine Dutch companies are member of just one 2-club: a social circle of size 10, which was maximally overlapped (apart from one firm) by a single hamlet of 12 companies. We chose this hamlet as the Dutch pivot, consisting of 10 Dutch and two German companies (See Figure 10).
From Table 2 we see that this Dutch pivot has a scope of 26.1%, similar to that of the German regional pivot.

Finally, for the three Scandinavian members of the European Union we found mixed configurations.

*Sweden.* For Sweden we found a single social circle of 15 companies (12 Swedish, one Norwegian, one Swiss and one British). As there were no larger hamlets with maximal overlap we selected this social circle as a pivot of the Swedish regional network. This Swedish regional pivot, less widely meshed then a hamlet, was narrowly organized around three central pairs, all from from *Electrolux AB* with *Ericson AB*, *Volvo AB* en *Svenska Cellulosa AB,* repsctively (See Figure 11).
Its 15 companies (12 Swedish, one Norwegian, one Swiss, one British) have a scope of 419 2-clubs of the European corporate borough: each of those 2-clubs has at least one company of that social circle as a member. Together these 419 2-clubs form 19.7% of the 2128 2-clubs of the borough.,.

*Denmark*. For the Danish region we could only find as a pivot an almostminimal hamlet of size 7, consisting of two Danish, two Swedish, one Norwegian and two British companies (See Figure 12). Hence this hamlet is less a pivot for a Danish region, but more a regional Scandinavian one, supplemented with two British companies. Its seven companies have a modest scope of only 277 2-clubs, or 13.0% of the European corporate borough (See Table 2).

*Finland*. For all five Finnish companies we found a single common hamlet of size 8, consisting of these five Finnish companies, one German and two Dutch firms (See Figure 13). This mixed hamlet, containing all five Finnish companies was thus selected as a pivot for the Finnish regional network. Its eight companies have a scope of 449 or 21.1% 2-clubs of the European corporate borough: each of those 2-clubs has at least one company of that hamlet as a member. Together these 449 2-clubs form 21.1% of the 2128 2-clubs of the borough.

## Non European Union

At the bottom Table 2 also contains data for two countries outside the European Union: Norway and Switzerland.

*Norway*. Four Norwegian companies shared one common 2-club, a hamlet of size 7, with 3 Swedish companies (See 14). As a regional pivot this mixed bi-national hamlet (4 Norwegian, 3 Swedish), apparently is appended to the Swedish network.

Its 7 companies have a scope of only 33 2-clubs or 1.6% of the European corporate borough, the lowest in Table 2, suggesting an extremely marginal position in the close community structure of the European Borough.

*Switzerland*. Six Swiss and two German companies can be designated as a pivot of the Swiss regional corporate network (See Figure 15). Its 8 companies have a scope of 478 2-clubs or 22.5% of the European corporate borough (see Table 9).

## Summary

The technique of regional pivotal 2-clubs enabled us to select regional pivots for 12 European countries: 10 of them members of the European Union and 2 non-members. We defined their scope of the borough: the number or percentage of the (2128) 2-clubs of the borough which share at least one firm of a pivot. The scope of a pivot indicates its coverage of the area of close communication in the corporate network, as determined by the European borough. The high value of 76.0% for the scope of the large French pivot (size 22) appears to confirm the centrality of the French regional network in the close communication sub-network of that borough. The high scope of 71.5% for the Belgian pivot (size

14) reflects the inclusion of the francophone Belgian firms in the French regional network.

On the other hand the moderate scope (29.5%) of the German regional pivot, with size 21 the second largest pivot, suggest a more peripheral North European position in the European borough.

Similarly, the Italian regional pivot of half that size (10), with scope 37.2%, suggests a similar marginal South European position in the borough.

Most interesting, if not spectacular, is the position of the British regional pivot. With about the same size (11) it has the highest scope of all, 81.0% of the 2-clubs of the European borough, which suggests a much wider range of close communication across that borough.

On the other hand we note the isolated positions suggested by the regional pivots for Spain and Norway with almost minimal size (6 and 7, repectively) and scope: 3.4% and 1.6%.

These results raise the question of interregional aspects of these regional networks. In the next section we try to get an impression by means of 'interlocks' of regional pivots.

# 5   'Interlocking' regional pivots

In Table 2 we noted for some regional pivots, such as the French and German pivots, a homogeneously national composition, for others, *e.g.* the United Kingdom and Denmark the pivots had a more nationally mixed composition.

| REGION | INTERLOCKS PIVOTS** |
|---|---|
| *EU* | |
| France | UK (2), Italy (1), Belgium (4) |
| Germany | NL (1), Finland (1), Switzerland (3) |
| United Kingdom (UK) | France (2), Belgium (2), NL (2), Sweden (2), Denmark (3), Finland (1) |
| Italy | France (1), Spain (3) |
| Belgium | France(4), UK (2) |
| Netherlands (NL) | Germany (1), UK (2), Finland (1) |
| Sweden | UK (2), Denmark (3) |
| Denmark | UK (2), Sweden (3), Norway (1) |
| Finland | Germany (1), UK (1), NL (1) |
| Spain | Italy (3) |
| *Non EU* | |
| Norway | Denmark (1) |
| Switzerland | Germany (3) |

*Table 3: Interlock of regional pivots*

*\*\* Number of firms in overlap between parentheses*

This suggested the idea to investigate if, and to what degree these pivots 'interlock' pairwise. Two pivots interlock pairwise when they share at least one firm (node) of the  network, and the degree of interlock is given by the number of firms in their overlap.

The results are summarized in Table 3.  In the second column we see for each country the list of  other regional pivots, with which they share at least one firm, and between parentheses the number of firms shared.
By far the most interlocked pivot is that of the United Kingdom. It covers half the set of regional pivots: France, Belgium, The Netherlands, and the three Scandinavian EU member countries (Sweden, Denmark,and Finland). Moreover, except for that with Finland (1) most of these interlocks are multiple where the pivot for Denmark, with 3 shared firms appears to be particularly strongly attached to that of the UK.
Notably, there are no interlocks of the British regional pivot with those for Germany and Italy: the German pivot has a solely North European orientation with interlocks with the Dutch, Finnish and Swiss pivots, whereas the pivot for Italy marks a clear Latin European position, interlocking with the French and Spanish pivots only.

With each 3 common firms, the Swiss and Spanish regional pivots appear to be strongly tied to those of a neighbour: for Switzerland the German pivot and for Spain the pivot of its Mediterranean neighbour Italy.
For non EU member Norway the pivot, is almost isolated from the other 11 regional pivots, sharing just one firm with the mixed Scandinavian pivot of Denmark.

# 6  Discussion

We have reanalysed Heemskerk's [2] corporate network of  286 of the largest European companies for the year 2010, from the perspective of close communication areas as defined by its boroughs and 2-clubs, in particular its social circles and more widely meshed hamlets ( [4],  [5], and [1]).  Apart from three minimally small boroughs/2-clubs (sizes 3-4) this network contained one giant borough of 225 firms, covering 79% of all firms in the network and 87% of its dominant component.
With a diameter of 7 this European borough, containing virtually all close communication between companies  in the form of  2128 2-clubs, formed a widely stretched body of close communication in the European corporate network of 2010.
Analysis of the distribution and composition of the 2-clubs suggested a dense and central  dominance of French companies in the borough. For instance, the largest 2-club was a coterie with GDF Suez SA as centre, consisting of 21 francophone companies and 6  firms from 5 other countries. Moreover, the 35 largest social circles and the 44 largest hamlets contained predominantly French firms.
The nest largest 2-clubs consisted of German companies.
Because in a dense network  the multitude of 2-clubs are overlapping heavily, we used the concept of *regional pivot* a single 2-club shared by a set of  firms from a common region, which enabled us to determine its scope in the borough (percentage of 2-clubs of the borough sharing at least one firm with the pivot) and the interregional, *i.e.* inter-pivot structure.

Heemskerk's dataset for 2010 covered 16 European countries. Our European borough for 2010, did not contain any major company from Portugal or Greece. The firms from Luxembourg and Ireland were all part of the French, respectively British regional networks, and were not considered separately. Hence we analysed pivots for 12 European countries: 10 from the European Union and 2 from non-EU countries. The results confirm the central position of the French regional level with a scope covering 76% of the 2-clubs of the European borough, incorporating the francophone Belgian regional pivot with its scope of 71.5%. This central position was already foreshadowed by the largest 2-club GDF Suez, which has the largest scope of all: 84.8%.

The moderate scope (29.5%) of the German regional pivot, though with size 21 the second largest pivot, suggested a more peripheral, if not secluded, North European position in the European borough. A similar, peripheral South European position was seen for the Italian pivot.

Rather spectacular were the results for the British regional pivot: with half the size of the French and German ones, it has the highest scope, 81%, of all regional pivots, well distributed over the Continental part of the borough.

In an additional analysis we investigated the inter-pivot relations in terms of common firms ('pivot interlocks'). The results confirmed these impressions: by far the most interlocked pivot is that of the UK, covering half the set of regional pivots, but excluding those of Germany and Italy.

Also notable was the result for Spain: its pivot was just closely tied to Italy, but for the rest with a scope of 3.4% quite isolated.

In a similar way this concept of pivotal 2-clubs can be used for other types of companies, for instance type of industry. To do so is beyond the scope of this paper and one possible line of research.

Thus we hope to have shown that the framework and method of boroughs and 2-clubs can add new perspectives to the analysis of corporate and other social networks in addition to those offered by the available methods.

# 8 Appendix  Regional pivots

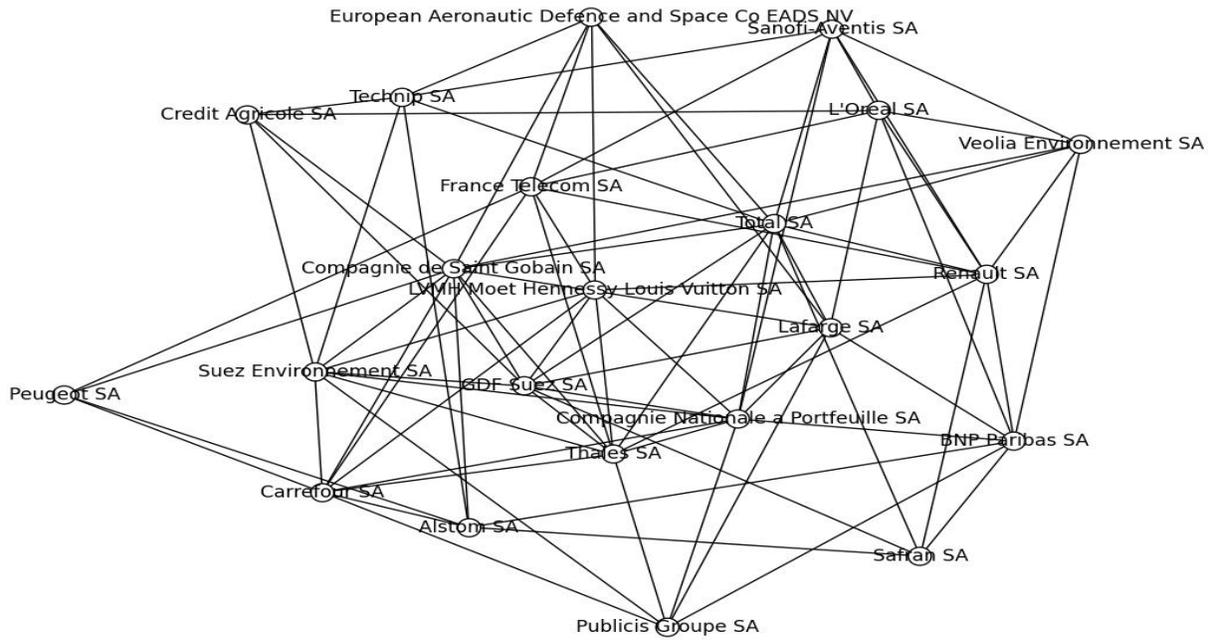

*Figure 4: Pivotal French Hamlet of 22 French companies*

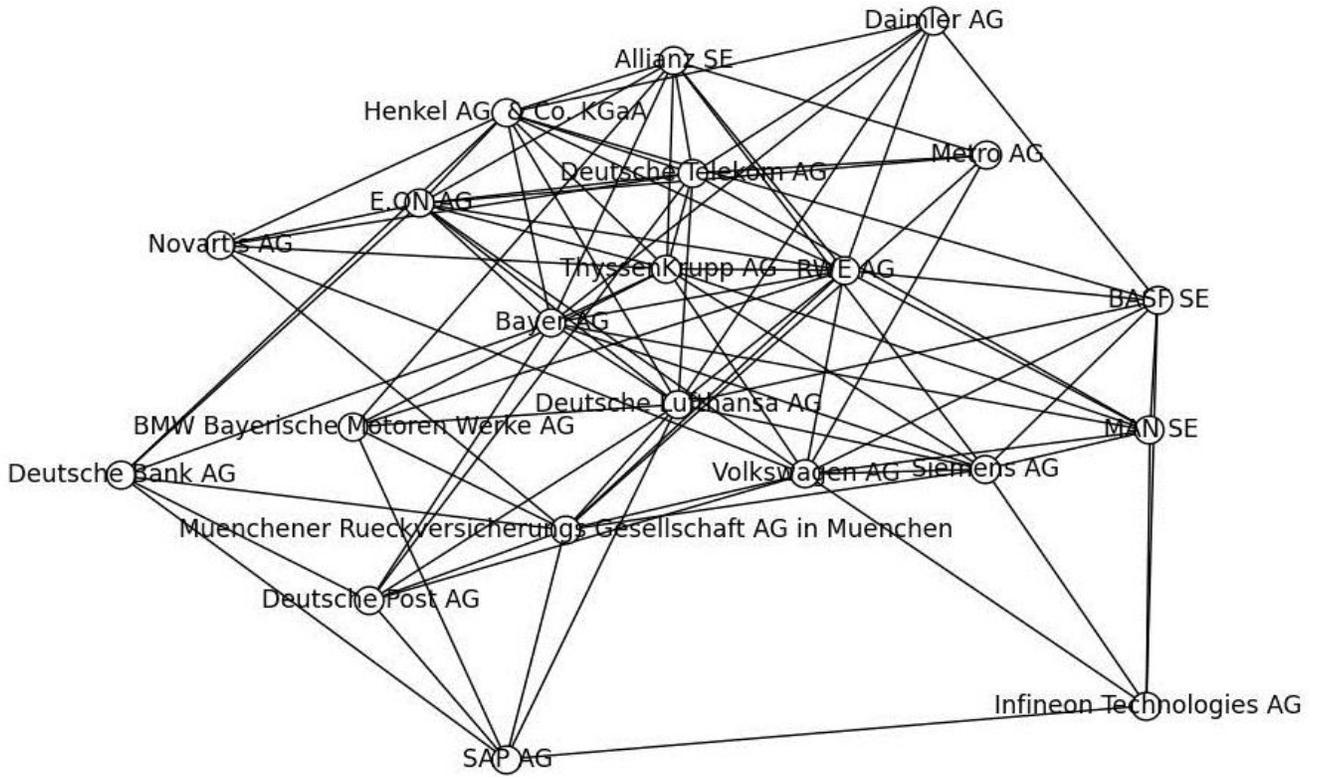

*Figure 5: German pivotal hamlet: 20 German companies and 1 Swiss*

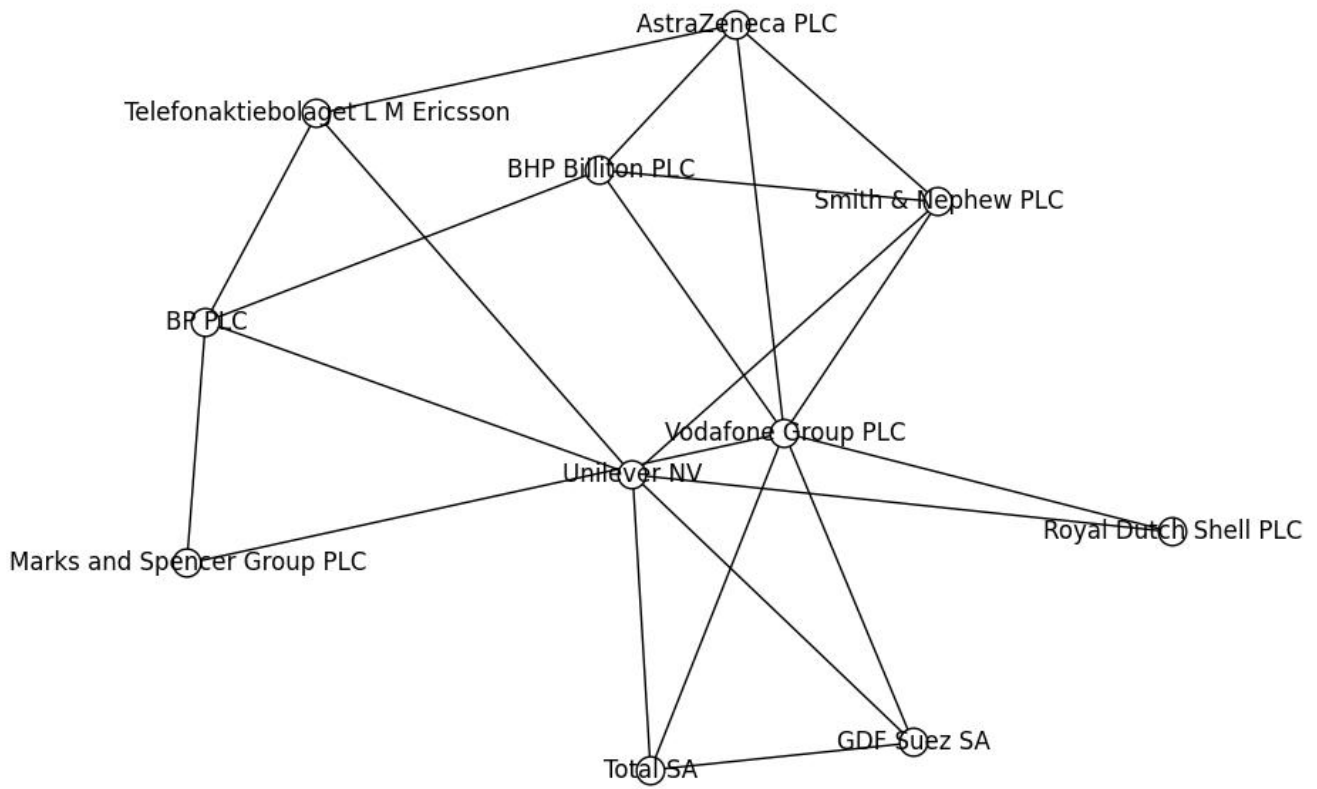

*Figure 6: British pivot. Hamlet: 7 British, 2 French, 1 Dutch, 1 Swedish firms*

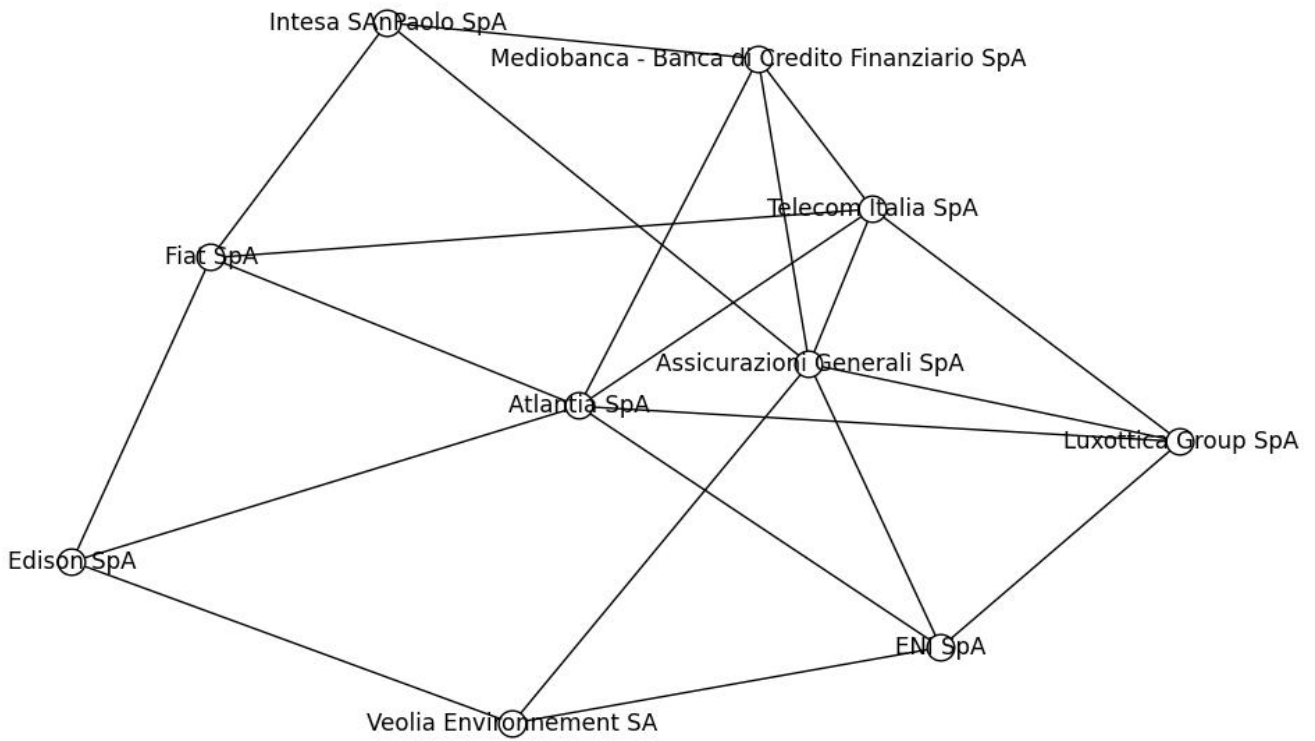

*Figure 7: Italian pivot. Hamlet: 9 Italian companies and 1 French*

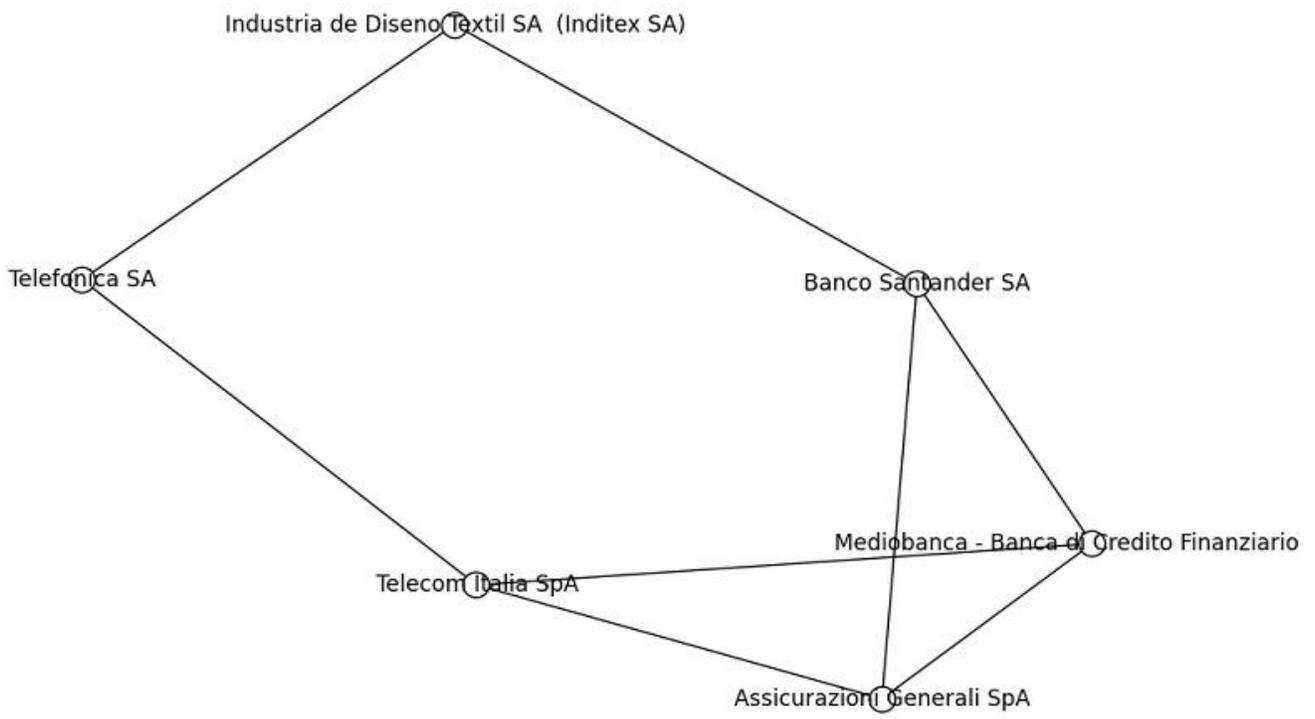

*Figure 8: Spanish pivot: Hamlet: 3 Spanish and 3 Italian firms*

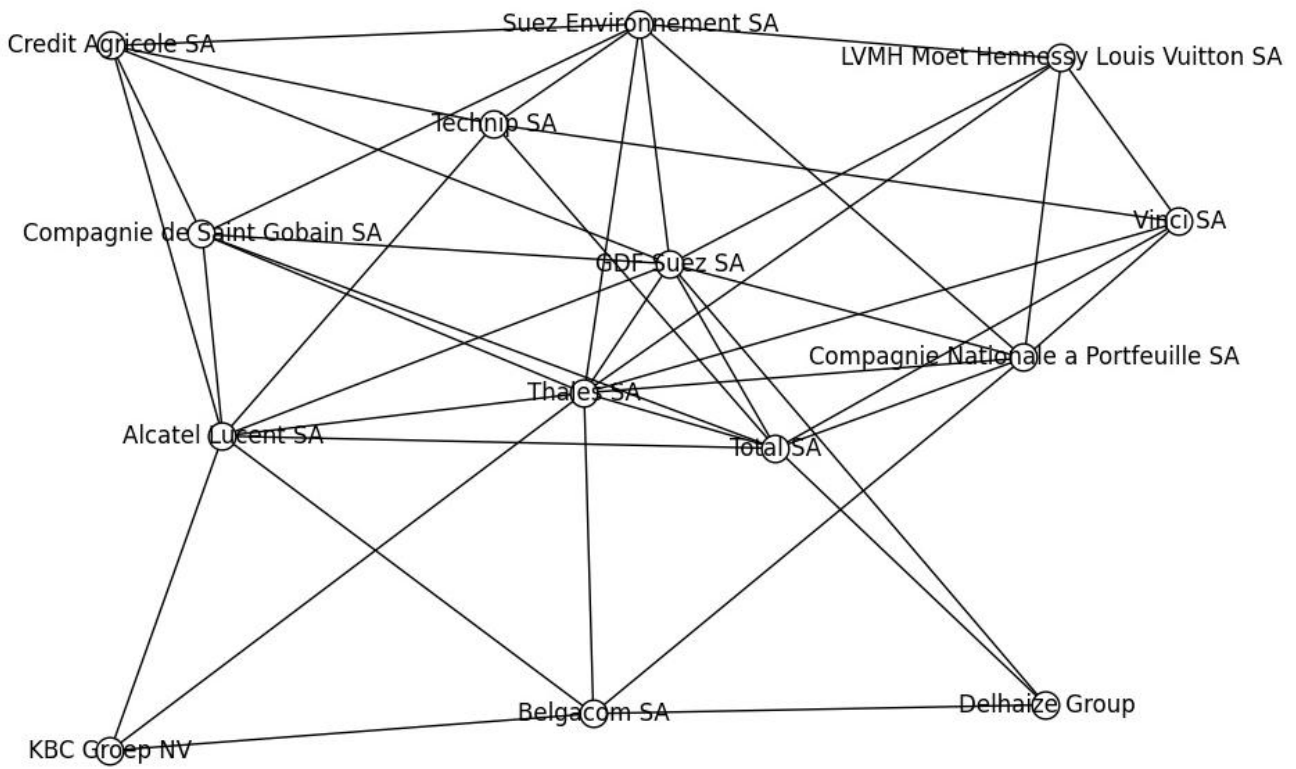

*Figure 9: Belgian pivot. Hamlet: 4 Belgian and 10 French firms*

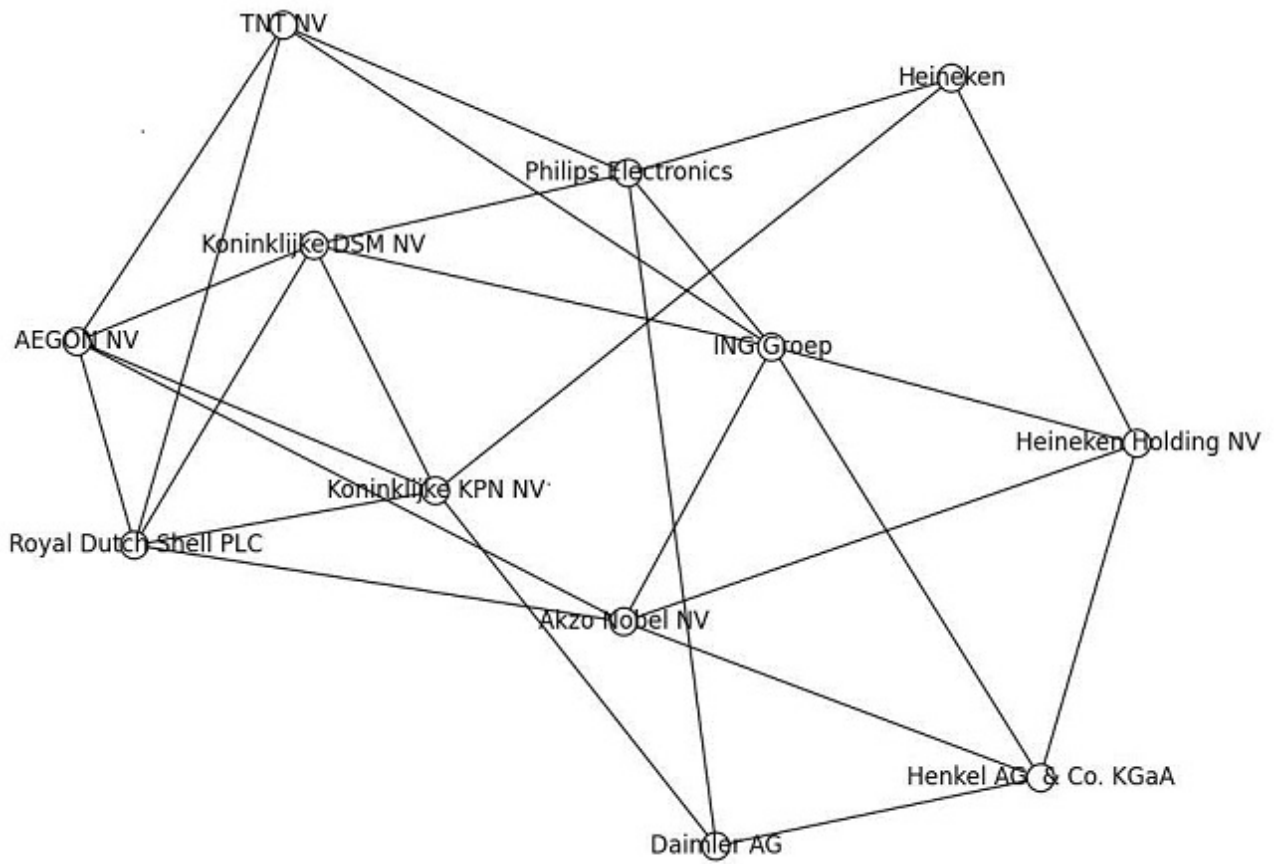

*Figure 10: Dutch pivot. 9 Dutch companies and 1 German*

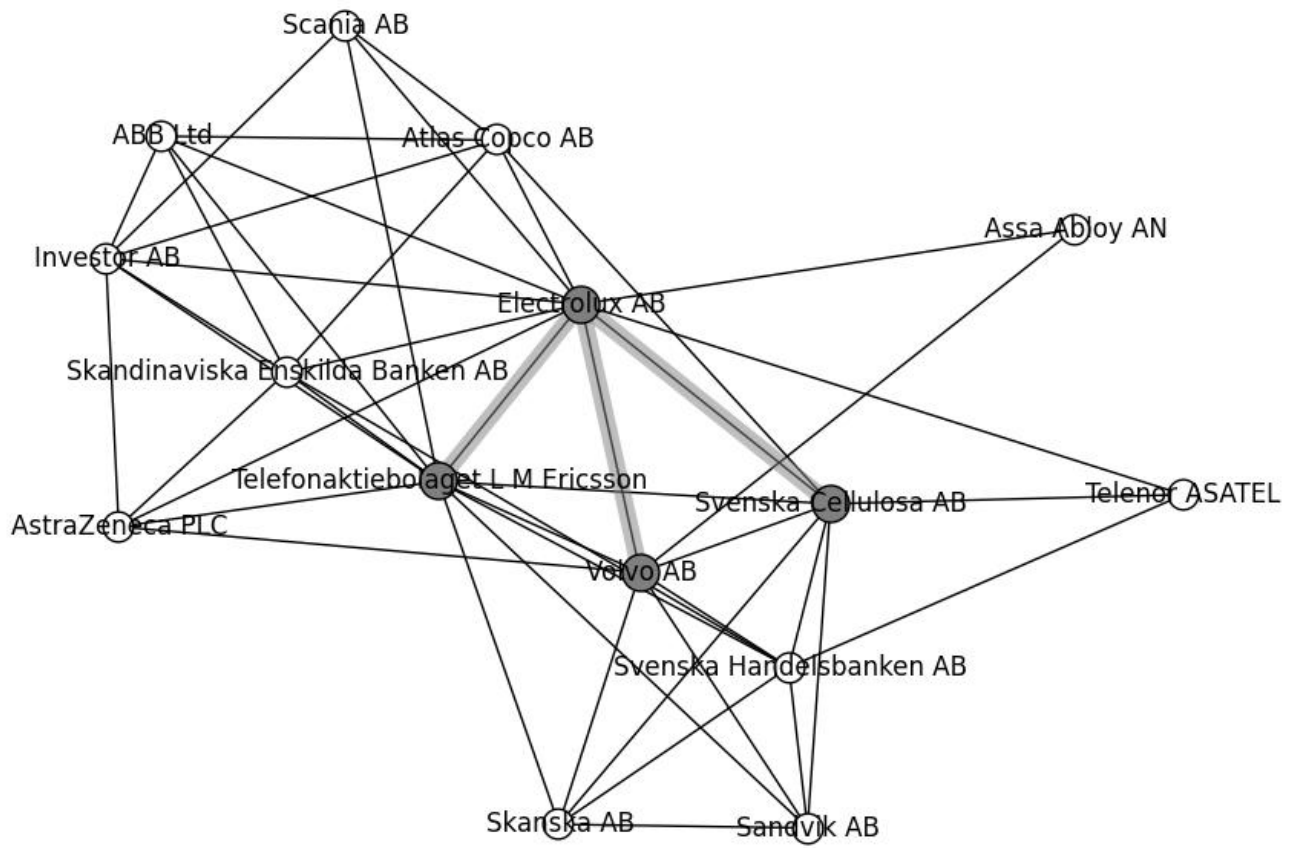

*Figure 11: Swedish pivot: Social circle with three central pairs.*
*12 Swedish, 1 Norwegian, 1 Swiss, 1 British firms*

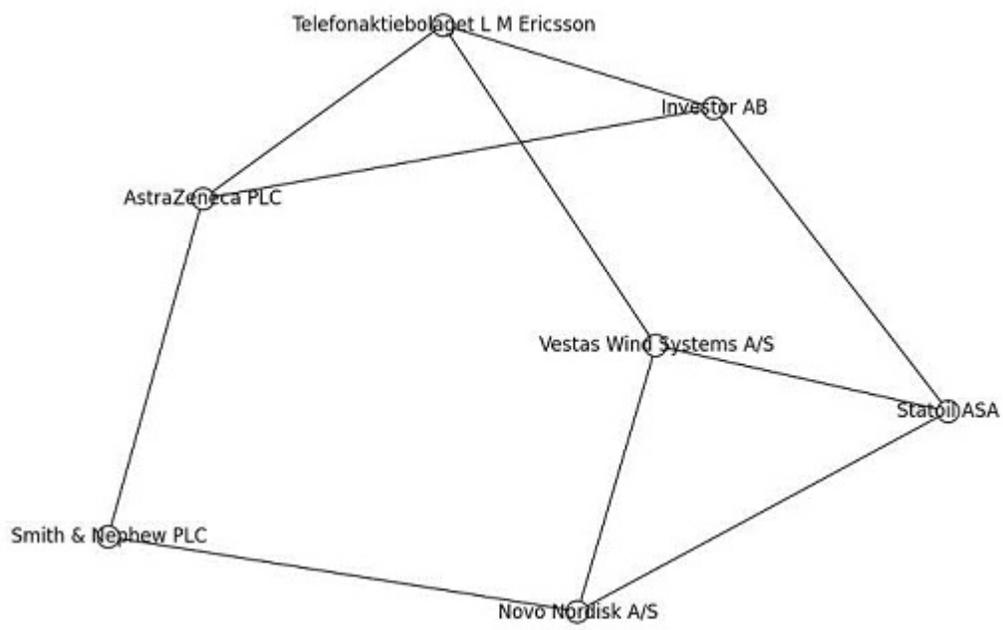

*Figure 12: Danish pivot. Hamlet: 2 Danish, 2 Swedish, 1 Norwegian and 2 British firms*

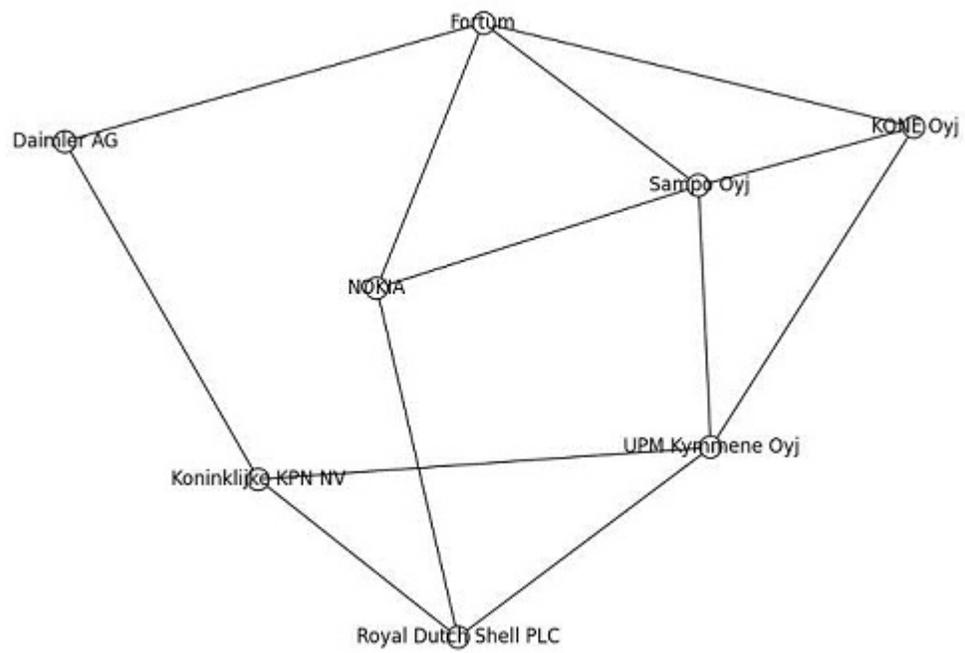

*Figure 13: Finnish pivot. Hamlet: 5 Finnish, 2 Dutch, 1 German firms*

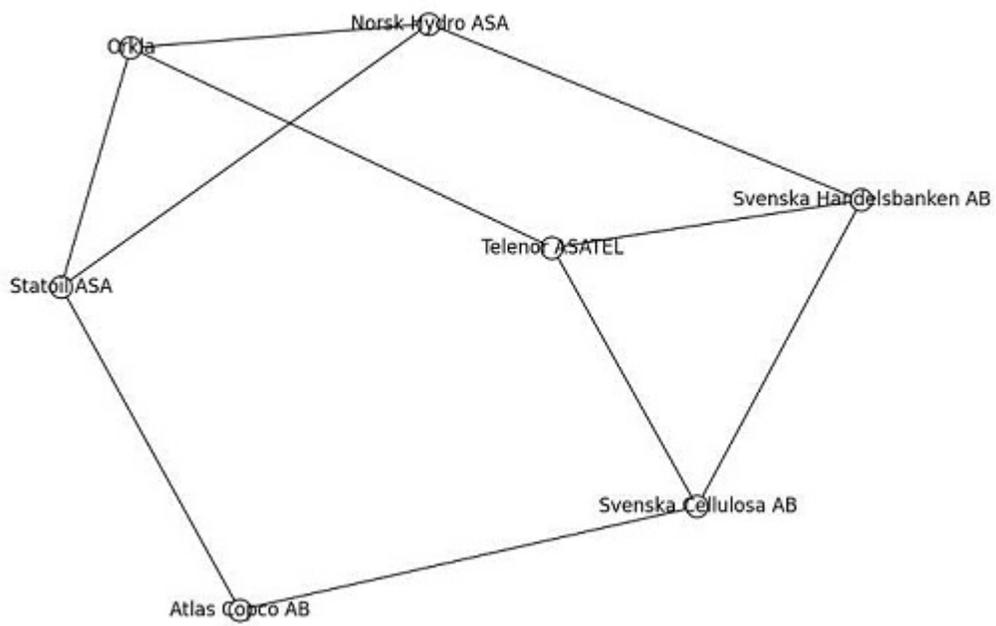

*Figure 14: Norwegian pivot. Hamlet: 4 Norwegian and 3 Swedish companies*

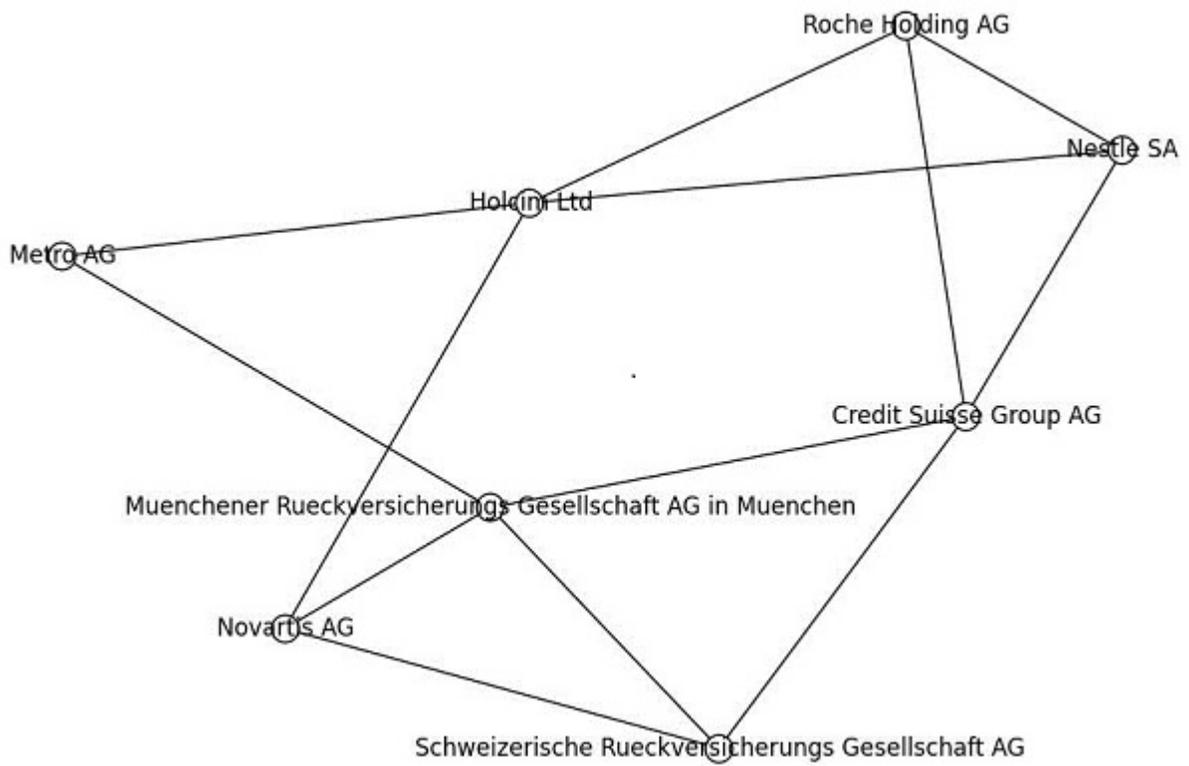

*Illustration 15: Swiss pivot. Hamlet: 6 Swiss, 2 German companies*